\documentclass[manuscript]{acmart}

\usepackage{subfigure}
\usepackage{epigraph} 
\usepackage{graphicx}
\usepackage{tabularx}
\usepackage{array}
\usepackage{pdflscape}
\usepackage{multirow}
\AtBeginDocument{%
  \providecommand\BibTeX{{%
    \normalfont B\kern-0.5em{\scshape i\kern-0.25em b}\kern-0.8em\TeX}}}

\setcopyright{rightsretained}
\copyrightyear{2023}
\acmYear{2023}
\acmDOI{XXXXXXX.XXXXXXX}

\acmPrice{15.00}
\acmISBN{978-1-4503-XXXX-X/18/06}

\begin{document}

\title{Scaffolding Ethics-Focused Methods for Practice Resonance}


\author{Colin M. Gray}
\email{gray42@purdue.edu}
\affiliation{%
  \institution{Purdue University}
  \streetaddress{401 N Grant Street}
  \city{West Lafayette}
  \state{Indiana}
  \country{USA}
  \postcode{47907}
}
\author{Shruthi Sai Chivukula}
\email{schivuku@iu.edu}
\affiliation{%
  \institution{Indiana University}
  \streetaddress{901 E 10th St Informatics West}
  \city{Bloomington}
  \state{Indiana}
  \country{USA}}

\author{Thomas Carlock}
\email{tcarloc@purdue.edu}
\author{Ziqing Li}
\email{li3242@purdue.edu}
\affiliation{%
  \institution{Purdue University}
  \streetaddress{401 N Grant Street}
  \city{West Lafayette}
  \state{Indiana}
  \postcode{47907}
  \country{USA}
}

\author{Ja-Nae Duane}
\email{jduane@bentley.edu}
\affiliation{%
  \institution {Bentley University, Departments of Information and Process Management}
  \streetaddress{175 Forest St.}
  \city{Waltham}
  \state{Massachusetts}
  \country{USA}}  

\renewcommand{\shortauthors}{Gray, et al.}

\begin{abstract}
  Numerous methods and tools have been proposed to motivate or support ethical awareness in design practice. However, many existing resources are not easily discoverable by practitioners. One reason being that they are framed using language that is not immediately accessible or resonant with the felt complexity of everyday practice. In this paper, we propose a set of empirically-supported ``intentions'' to frame practitioners' selection of relevant ethics-focused methods based on interviews with practitioners from a range of technology and design professions, and then leverage these intentions in the design and iterative evaluation of a website that allows practitioners to identify supports for ethics-focused action in their work context. Building on these findings, we propose a set of heuristics to evaluate the practice resonance of resources to support ethics-focused practice, laying the groundwork for increased ecological resonance of ethics-focused methods and method selection tools. 
\end{abstract}

\ccsdesc[500]{Human-centered computing~Empirical studies in HCI}
\ccsdesc[500]{Human-centered computing~HCI design and evaluation methods}
\ccsdesc[500]{Human-centered computing~Usability testing}

\keywords{design methods, design and technology practice, ethics-focused methods, ethics, scaffolding}


\maketitle

{\color{red}\textbf{Draft: September 13, 2022}}

\section{Introduction}

Supporting ethically-focused design practices has long been a goal in the HCI community, with various attention over the past three decades towards the development of a meaningful code of ethics~\cite{Wolf2019-nm,Gotterbarn1998-la,Gotterbarn2017-nw}, the identification of accreditation or programmatic requirements in technology education that address ethical responsibility~\cite{Bucciarelli2008-zc,Fiesler2020-vo,G_Pillai2021-hm}, and an increasing body of methods, toolkits, and other resources that are intended to encourage ethically-focused or ethically-sensitive design practices~\cite{Friedman2017-rd,Shilton2018-ro,Chivukula2021-xk}. 

Ethical dimensions of practice are known to be complex, contingent, and situated in relation to a wide range of factors, which include the ethical knowledge and judgments of individual practitioners~\cite{Chivukula2021-oj,Debs2022-dq,Watkins2020-zr,Dindler2022-ny}; the existence of standards, resources, and processes that support ethics-focused inquiry~\cite{Gogoll2021-lq,DIgnazio2020-is,Vermaas2015-vp,VSDbook}; and the mediation of organizational and practitioner forces to encourage action~\cite{Gray2019-wa,valuelevers,Wong2021-pv,Chivukula2021-oj,Pillai2022-xy,Chivukula2020-oy,Dindler2022-ny}. Historic work on supporting ethical awareness has included the development of a range of methods to support ethically-focused action, including activities to design for privacy \cite{Wong2017-wt, Shilton2020-uk}, acknowledge data security \cite{reshape-paper}, highlight and correct gender representations in software \cite{Oleson2018-ml}, and implement value-sensitive practices~\cite{Friedman2017-rd,VSDbook}, among numerous others. A recent collection of such methods by Chivukula et al.~\cite{Chivukula2021-xk} has described a set of 63 such methods with an explicit ethical focus, including 15 methods drawing from the academically-popular \textit{Value Sensitive Design} approach as well as 14 additional methods that are part of the \textit{Design Ethically} or \textit{Ethics for Designers} toolkits created by practitioners, which have not previously been addressed in HCI literature. While resources clearly \textit{exist} to support ethically-focused work, relatively little attention has been paid to \textit{how these methods are discovered, compared, and used by designers}---what we describe as the \textit{scaffolding} that encourages and supports practitioner engagement in method selection and use. Additionally, while scholars have described the need for design knowledge and tools to be \textit{resonant} with the ecological complexity of design practice, few methods have been validated for these resonance qualities in relation to specific use contexts or practitioner goals~\cite{Stolterman2008-ho,Stolterman2008-ty,Gray2022-na,Gray2014-fk}. In addressing the scaffolding of ethics-focused methods in practice-resonant ways, we build upon the historic interest of HCI as a discipline in creating tools that scaffold user behaviors, from early examples such as Carroll's ``training wheels'' approach to software learning in the 1980s~\cite{Carroll1984-vx} to modern ``onboarding'' practices on websites and apps that increase learnability and support ``quick wins'' on the part of users~\cite{Strahm2018-ep}. Using scaffolding as a point of focus we ask: How can design and technology practitioners leverage resources to build a stronger sense of ethical awareness and ability to act? Scaffolding these behaviors to result in meaningful connections between existing resources and ethically-focused practices is the core issue we seek to address in this paper.

In this paper, we use an action research (AR) approach~\cite{Hayes2014-mp} to identify new means of supporting technology and design practitioners' engagement in ethical design complexity, including means of better supporting ethical awareness and action in their everyday design practices. Our approach includes three parts: 1) identification of intentions that drive ethical action; 2) the creation of an ethics-focused method discovery tool; and 3) the iterative evaluation of this tool that resulted in a preliminary set of practice-resonance heuristics. First, we conducted a new analysis of interviews previously conducted with 25 technology and design practitioners, which we used to create a set of seven ``intentions'' that practitioners can use to frame their pursuit of new methods to support ethically-engaged work. Second, building on these intentions, we designed an interactive method discovery website that contained an existing set of 63 ethics-focused methods proposed by Chivukula et al.~\cite{Chivukula2021-xk}, which we then tagged with intentions and other metadata to support practitioner identification and use of ethics-focused methods. Third, we iteratively evaluated the method website with 10 design practitioners and students, supporting improvements to the website itself and also aiding us in identifying an initial set of six heuristics that support the creation of practice-resonant scaffolds for ethically-focused design work. In this multi-phase action research project, our guiding research questions were as follows:

\begin{enumerate}
    \item What intentions can be used to frame a designers desire to support their ethics-focused practice in ecologically meaningful terms? 
    \item How can we scaffold the selection of methods in ways that support designers' ethical awareness and action? 
    \item What qualities of method discovery appear to increase practice resonance? 
\end{enumerate}

Our contribution in this paper is three-fold: First, we describe a set of empirically-supported ``intentions'' that frame practitioners' motivations for selecting a method or tool to support ethically-focused action, laying the groundwork for the targeted creation of new methods and more effective scaffolding of existing practices. Second, we propose a means of employing intentions to structure practitioners' selection of methods that are relevant to support ethically-focused practice, providing an example of how instrumental judgments and methods might be connected to support emergent practitioner needs. Third, we identify an initial set of heuristics to evaluate resources for practice resonance, laying the groundwork for the creation and iteration of methods and other tools in ways that accounts for the ecological complexity of real-world practice.


\section{Background Work}

\subsection{Methods and Their Use by Practitioners}

Designers rely on a range of sources of knowledge to support their everyday work \cite{Nelson2012-ov,Lawson2004-wf,Lowgren1999-oy}, including theory, practical design exemplars, and many other types of intermediate-level knowledge in a space that L\"{o}wgren describes as ``non-empty'' \cite{Lowgren2013-jt}. In this paper, we specifically focus on one form of this intermediate-level knowledge, the \textit{design method,} which contains the potential for design action (Gray~\cite{Gray2016-tt,Gray2022-na} refers to this element as the method ``script'') but does not fully define how, or in what form, that potential might take when the method is performed \cite{Lowgren1999-oy,Rittel1984-ga,Gray2016-ux,Goodman2011-ak,Goodman2013-bb}. 

When engaging the complexity of real world practice, designers must select appropriate forms of knowledge that will support and move their design work forward, relying upon their capacity for design judgment \cite{Nelson2012-ov,Stolterman2008-ho,Goodman2011-ak}---and in relation to the use of methods particularly, their use of instrumental judgments inform which method is selected, with what intention, and how the designer knows when the method has been efficacious (or not) \cite{Gray2018-cf,Murdoch-Kitt2020-sw,Gray2016-ux}. In previous studies, practitioners report that methods use is, in a sense, is more about ``mindset'' than the method itself~\cite{Gray2016-ux}, while also underscoring that methods can be important ``mental tools'' that can structure the complexity of practitioners' everyday work~\cite{Daalhuizen2014-cr}. In the current state of the literature, a somewhat uneasy relationship is present between codification and performance, with some scholars such as Goodman focusing on the improvisatory qualities of methods use in practice \cite{Goodman2013-bb} while other scholars such as Daalhuizen and Cash \cite{Daalhuizen2021-qa} focus on how elements of methods may predict future patterns of performance. For the purposes of this paper, we recognize that method codification and performance are connected, but are often considered separately~\cite{Gray2022-na}. 
Many scholarly efforts in relation to design methods have focused on the validation of individual methods, often in relation to specific lab-managed protocols or other means of controlling for design complexity, including both experimental and implementation-focused studies. Eisenmann et al.~\cite{Eisenmann2021-ct} report on a recent survey of such design validation literature, describing issues of alignment that relate to a lack of shared vocabulary across methods (including relevant metrics to compare performance), describing this challenge as follows: ``most design method developers set goals
for their own methods and therefore develop a separate operationalisation resulting in a multitude of metrics. This makes it very challenging to compare methods with each other and hinders researchers to build a common standard for similar methods.'' \cite[p. 633]{Eisenmann2021-ct}. As a more ecologically-focused metric of methods use, Gericke~\cite{Gericke2020-fb} describes the potential for individual methods to exist as part of a methods \textit{ecosystem}, ``provid[ing] numerous [\ldots] methods for different purposes and ways to combine and supplement them.'' 

In parallel with scholarly inquiry into method validation, over the past decade, a range of collections of methods have been published to support design students, practitioners, and non-design professionals alike. Popular print-based methods collections such as \textit{Universal Methods of Design}~\cite{Hanington2019-jd} and the \textit{Delft Design Guide}~\cite{Van_Boeijen2014-vp} have become common tools in design education settings, while other digital collections such as a collection of methods as part of IDEO's \textit{Design Kit}~\footnote{\url{https://www.designkit.org/methods}} are intended to support design thinking-focused work in a range of educational and practitioner settings.

In this paper, our focus is practice-led and action oriented~\cite{Gray2014-fk,Colusso2019-mw}, seeking to result in ``cycle-around'' research that connects practitioner-led bubbling up of experiences relating to ethical support with the creation of a tool that can then enter the practical lexicon of everyday technology and design work. Thus, we engage with historic framings of methods in a scholarly sense while also acknowledging the growing popularity of collections of methods which allow practitioners to navigate among and select methods that add potential value to their practice. To make methods tractable for our analysis and design work, we rely upon the analysis of a collection of ethics-focused methods described in a recent survey by Chivukula et al.~\cite{Chivukula2021-xk} as well as definitional work around components of methods by Gray~\cite{Gray2022-na} and other related design theory scholarship~\cite{Lowgren1999-oy,Harrison2006-dr,Goodman2011-ak}. For instance, to inform our mapping of methods in relation to intentions in Section~\ref{sec:intentions}, we leverage Gray's concept of the method \textit{core}, which describes the ``central conceit or framing metaphor that makes the entire method, or a portion of the method, coherent and potentially interchangeable''~\cite[p. 8]{Gray2022-na}. 


\subsection{Supports for Ethically-Focused Practice}

Scholars have increasingly paid attention to the role that ethics plays in describing normative dimensions of design practices \cite{Shilton2018-ro,VSDbook}, including conversations relating the designer’s goal in producing socially and ethically responsible design outcomes~\cite{valuelevers,Steen2015-mt,DesignFictionMemos,Shilton2018-sw}, the identification of relevant ethical standards to support appropriate design behaviors~\cite{Gogoll2021-lq,Gotterbarn2017-nw}, and the mapping of stakeholders that hold responsibility for the ethical soundness of design outcomes~\cite{Wong2021-pv,Gray2019-wa,Chivukula2021-oj,valuelevers,Gray2018-bh}. Over the past two decades, methods and approaches that have an ethical focus have been developed in academic literature and by design practitioners and design organizations~\cite{Friedman2017-rd,Shilton2018-ro,Chivukula2021-xk,Vermaas2015-vp}. For example, value-sensitive methods developed and evaluated as part of the \textit{Value Sensitive Design} approach have been proposed and disseminated, often using academic publications as a primary vehicle for evaluation and publication\footnote{See a list of these methods and related publications at \url{https://vsdesign.org/vsd/} and \cite{Friedman2017-rd}}. In parallel, organizations and design practitioners alike have begun to create methods to support ethical reasoning in practice, including both expansive toolkits (e.g., Microsoft's Inclusive Design methodology) and sets of methods (e.g., practitioner Kat Zhou's \textit{Design Ethically} toolkit or Jet Gispen's \textit{Ethics for Designers} toolkit)~\cite{Chivukula2021-xk}. In total, Chivukula et al.~\cite{Chivukula2021-xk} have identified 63 such methods, including a mix of practitioner- and academically-produced tools with a wide range of foci, means of presentation, and potential opportunities for incorporation into design processes. Due to the diversity of these method authors and purposes, the current adoption and awareness of these support tools is mixed, with practitioner tools often not generating awareness by design scholars or educators and scholarly tools often not generating awareness by practitioners---a classic example of the research-practice gap that has been proposed by other HCI scholars~\cite{Colusso2019-mw,Brier2017-nl,Gray2014-fk}. With the recognition of this gap between academia and practice, we seek to prioritize a growing interest and body of work---largely driven by practitioners---that privileges practice-driven approaches to formulate guidance that is presented in ways that are resonant with the needs and requirements of practitioners~(e.g., \cite{ethicaldesignscorecards,whitehat,Monteiro2019-tj}).

In addressing this complex and rapidly evolving space, in this paper we rely upon previous descriptions of factors that impact the adoption of ethics-focused approaches in practice. We ground our understanding of these practice contexts through Gray and Chivukula's description of felt \textit{ethical design complexity}, defined as: ``the complex and choreographed arrangements of ethical considerations that are continuously mediated by the designer through the lens of their organization, individual practices, and ethical frameworks''~\cite{Gray2019-wa}, which builds on a number of critical accounts of ethics in technology practice by Shilton, Wong, Steen, and others~\cite{Steen2015-mt,valuelevers,Shilton2018-sw,Wong2021-pv}. We also identified the importance of considering both the normative thrust of ethical guidance and the agency of the designer, building upon the work of Donia and Shaw~\cite{Donia2021-md} to take seriously the role of the practitioner. In depicting the role of practitioners in these complex settings, we take on a practice-led framing, seeking not to mandate specific forms of engagement with methods or resources that have been designed or otherwise asserted by scholars, but rather to better understand what kinds of supports or scaffolds are needed for practitioners to make reasoned judgments about their methods selection and use that may then inform their work practices.



\section{Our Approach}
In this paper, we use an action research (AR) approach~\cite{Hayes2014-mp} to engage with the interest of a range of technology and design practitioners in becoming more ethically aware and active in their workplaces, using both practical and emancipatory qualities of AR to enable and empower practitioners to shape their local work contexts with new ethical supports. We began our multi-phase project by leveraging interviews we conducted with 25 practitioners which elucidated their experiences addressing ethical dilemmas in their everyday work and ecological challenges they faced in confronting these challenges. Analysis of these interviews allowed us to frame a set of potential intentions for ethical support, grounded in practice contexts and vocabulary (detailed in Sections \ref{stage1}). Based on these intentions and prior work that has described a landscape of ethics-focused methods which, if adopted by practitioners, could increase the potential for ethical awareness and action~\cite{Chivukula2021-xk}, we designed a resource discovery site as a scaffold to support the ethically-focused practice of technology and design practitioners, students, and educators. Finally, we conducted an evaluation of this website to identify opportunities for improvement, characterizing both usability concerns and emergent heuristics for practice-resonance for this website and future ethics-focused resources.

\subsection{Stage 1: Creating Method ``Intentions''} \label{stage1}
We used a reflexive thematic analysis approach~\cite{Braun2021-dt} to characterize a set of framings, intentions, or filters that would illustrate various dimensions of support that practitioners may utilize to support their ethics-focused practice, importantly using vocabulary that was accessible and resonant with their everyday work. We iteratively and reflexively framed ``intentions'' through four rounds of analysis, resulting in a final set of seven intentions that we report in Section~\ref{sec:intentions}. In this stage, we drew on two sources of data. First, we analyzed a set of previously conducted \textit{interviews} with 25 practitioners that provided a robust description of ethical dilemmas and related complexity. These interviews were conducted with technology and design practitioners from a range of professional roles (e.g., UX designer, software engineer, data scientist, product manager) and were 60--90 minutes in duration. Each semi-structured interview included detailed discussions of ethical dilemmas the practitioner had faced, their awareness of the ethical dimensions of their practice, and strategies they had used to support ethically-informed action. Building on this overall structure, we sought to identify what needs the practitioners had in supporting their ethical awareness and responsibility in their everyday work, ecological factors that impact different forms of ethical mediation and complexity, and the disciplinary values engaged by different practitioner roles in an organization. This helped us frame the intentions to align practice-resonance and knowledge as learnt from the practitioners.
Second, we built upon descriptions of a \textit{collection of ethics-focused methods} and related vocabulary previously described by Chivukula et al.~\cite{Chivukula2021-xk}. This collection included an evaluation of 63 ethics-focused methods and an initial set of descriptors that described the ethical focus or ``core'' of each method. This collection enabled us to map the desire for ethical supports in our interviews with pragmatic supports that had already been proposed in published methods.

\begin{itemize}
  
\item \textit{Mapping Possibilities to Filter Methods:} In the first stage, we attempted to identify all of the possibilities of filtering the set of supports, tools, situations, or methods for ethics-focused practice based on the collection described by Chivukula et al.~\cite{Chivukula2021-xk}. Drawing from this existing content analysis of 63 ethics-focused methods through a set of derived descriptors, we built on the two proposed types of ``cores'' of a method, which included: 1) \textit{Value-focused cores}, such as eliciting values, critically engaging, and defamiliarizing; and 2) \textit{Action-focused cores,} such as framing, generating, evaluating, and consensus building. As a research team, we listed various permutations and combinations of the three value-focused and four action-focused cores. For example, ``\textit{Framing} + \textit{Eliciting Values} leads to Methods for framing a value-focused problem space,'' ``\textit{Generating} + \textit{De-familiarizing} leads to Methods for conceptualizing pluralistic concepts that were not considered before,'' and so on. This stage sensitized us to different ways that the cores of methods might relate to the needs of practitioners expressed in our initial interviews.

\item \textit{Making Intentions Colloquial:} In the second stage, we iterated on vocabulary that felt colloquial and ecologically-meaningful to technology and design practitioners, seeking to identify ways of presenting the mapped cores to make the intentions more accessible to practitioners in their everyday work. For example, instead of ``Framing + Eliciting Values,'' a more practice-resonant way of presenting the intention could be ``\textit{Reconsider Values in Relation to a Design Problem.}''

\item \textit{Converging on a Set of Practice-Resonant Intentions:} In the third stage, we curated the set of potential filtering possibilities with different types of vocabulary to prioritize among the wide range of intentions we derived from various combinations. This was informed through additional analysis of our interviews with technology and design practitioners in conjunction with other empirical accounts of ethical complexity in action~\cite{Gray2019-wa,Chivukula2021-oj,Boyd2021-sv,Wong2021-pv}, we identified recurrent needs for specific ethical supports, such as practitioners saying: ``\textit{I need strength to speak up to my team about values},'' ``\textit{I need to navigate through gray areas},'' ``\textit{I need to align the team when it comes to difficult decisions},'' and ``\textit{My team wants to foster privacy driven solutions},'' among others. We then mapped these ``naturalistic'' intentions from our data back to the cores and possibilities we identified in the first two stages to converge on the most common framings for intentions that could support ethics-focused practice. 

\item \textit{Drawing Independent Intentions}: In the fourth stage, we began to build intentions that were independent of each other, starting with a list of potential permutations and empirically-validated intentions from our own and others' empirical investigation of supports for ethics-focused practice. This iterative construction resulted in seven distinct intentions, which we list and detail in Section~\ref{sec:intentions}. We took inspiration from the range of ``outputs'' indicated by the methods collection proposed by Chivukula et al.~\cite{Chivukula2021-xk}, framing the intention phrases using an ``I want to...'' structure that a practitioner could easily connect with when they are looking to support, improve, develop, or leverage their ethical awareness, action, and/or responsibility.

\item \textit{Coding Methods under Intentions:} In the final stage, we non-exclusively coded the complete set of ethics-focused methods proposed by Chivukula et al.~\cite{Chivukula2021-xk} based on the seven drafted intentions by directly comparing the ``I want to\ldots'' intention to set of related descriptors from the dataset, including the ``cores'' and ``outputs'' of the method. We also drafted a one-sentence intention-focused summary for each method we evaluated, seeking to make the connection between the intention and method core visible and easily understood. The descriptors and short summaries were then added as part of a custom taxonomy on the website, which formed part of the resource scaffold that we further detail in Section~\ref{stage2}.
\end{itemize}

\subsection{Stage 2: Development of a Support Website} \label{stage2}
With this collection of methods and a framework of intentions in place, we focused the action research project on creating a tangible and pragmatic tool---a website that practitioners, students, and educators could use to empower and enable their own ethically-focused work. With the development of this site in mind, we had two key questions. 
\begin{enumerate}
    \item How do we introduce the collection of ethics-focused methods and what system of navigation would be most resonant with the goals practitioners have in their everyday design work?
    \item What parts of each individual method do we prioritize as necessary to introduce, and how do we format this set of information on method-specific pages in a way that would be useful and actionable?
\end{enumerate}
     
While we were able to draw inspiration from existing sites that map and facilitate exploration of methods in a general sense\footnote{E.g., \url{https://www.designmethodsfinder.com/}, \url{https://www.designkit.org/methods}, \url{https://best.berkeley.edu/best-research/thedesignexchange/}}, none of these sites were specifically focused on presenting methods with an explicit ethical valence, and importantly, no other site we located used vocabulary other than design phase (e.g., research, evaluating, framing, conceptualization), general design keywords (e.g., empathy, brainstorming, market, user), or logistical factors (e.g., material required, participants) as primary navigational criteria.

\begin{figure*}[!ht]
    \centering
    \includegraphics[width=\textwidth]{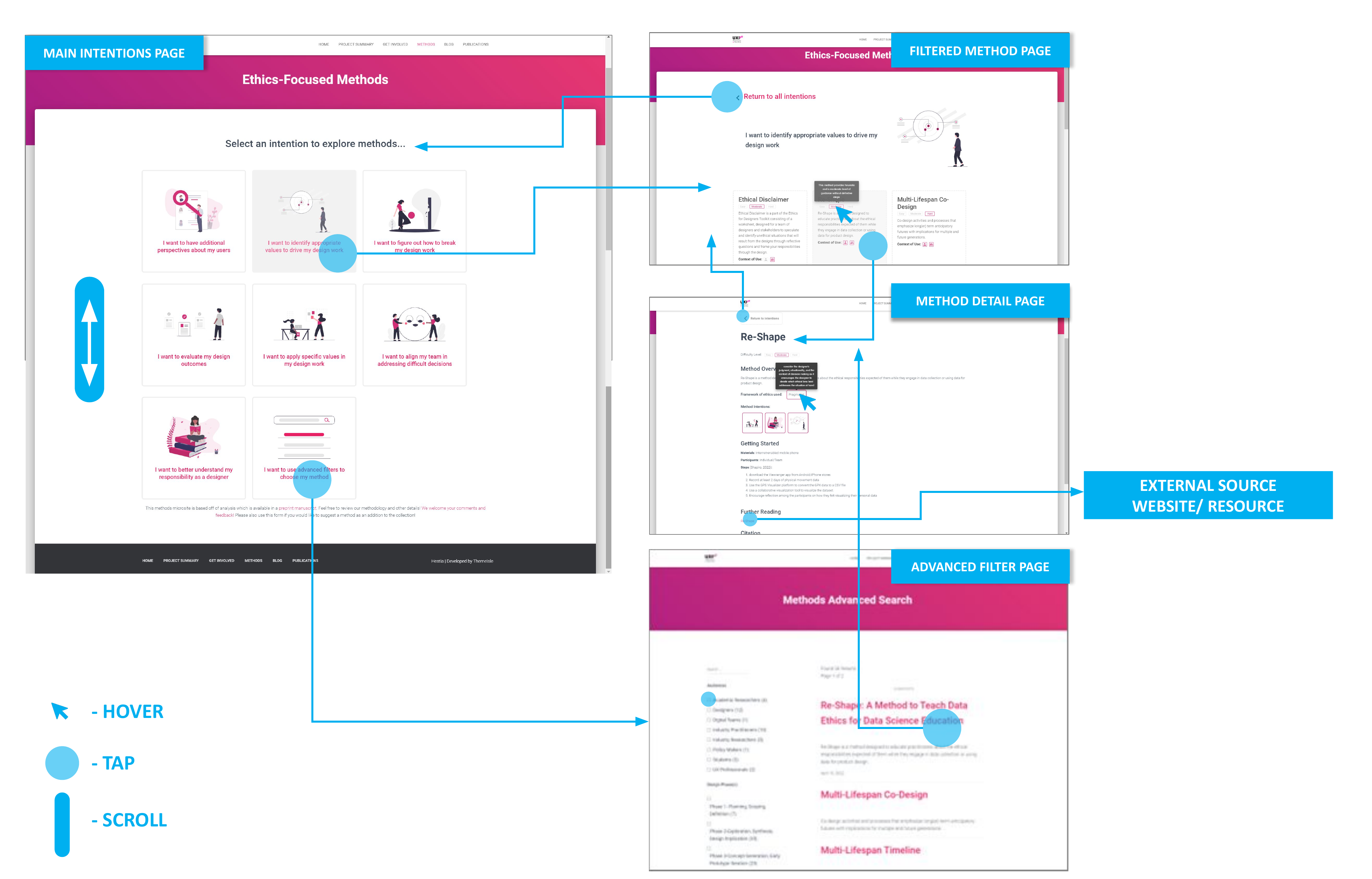}
    \caption{Initial interaction Flow for the intentions-focused navigation (left and top) and advanced filter navigation (bottom).}
    \label{fig:interactionflow}
    \Description[Interaction Flow with four images]{Interaction Flow with four images of different pages of the website. There are arrows linking clicked inputs to their connected navigation outcomes. The first image shows a page with panels depicting clickable intentions, the second image is a page of methods specific to the chosen intention, and the third image depicts an individual method page that results from selecting a method. The fourth image is a page with toggle-able filters that result in a list of relevant methods. Selecting a method on this page also results in navigation to the individual method page from the third image.}
\end{figure*}

We started our design work by focusing navigational attention on the method intentions we previously created. We wanted users to be able to view these intentions and identify one that resonated with one of their ``felt'' ethical motivations. Upon selection of a particular intention, they would then be taken to a page with all of the matching methods presented as clickable blocks, presented along with other relevant information. Upon clicking on one of the methods, then a detail view page of that particular method would be presented. This navigational approach followed a typical primary list->filtered list->detail view UI navigational pattern. 
We structured the various elements of these pages as follows:
\begin{itemize}
    \item On the \textit{primary list page}~(``Main Intentions Page'' in Figure~\ref{fig:interactionflow}, left), each intention was listed as a complete sentence (e.g., ``I want to align my team in addressing difficult decisions'') with an explanatory graphic.
    
    \item On the \textit{filtered list page}~(``Filtered Method Page'' in Figure~\ref{fig:interactionflow}, top right) of matching methods for a chosen intention, each method block included a short description that was written from the perspective of the intention selected along with a visual depiction of the difficulty of the method (easy, moderate, hard) and whether it was ideal for team or individual use.
    
    \item On the \textit{method detail page}~(``Method Detail Page'' in Figure~\ref{fig:interactionflow}, middle right), we presented method details in a structured way. The page started with a short \textit{Method Overview}, which was also used as the short description of the method in other parts of the site, such as the method blocks on the filtered list view. The next section was \textit{Getting Started}, containing any necessary materials for using the method (e.g., pens, whiteboards, templates), participants (e.g., team, individual), and steps drawn from the original method. Depending on the content of the method, the detailed information would either include actionable or outlined steps or a high level description of the principles or best practices that were informed by the method. The detail page also included a \textit{Citation} and/or \textit{Further Reading} link to other resources (e.g., a link to a template associated with the method), visual icons of all \textit{Method Intentions} that mapped to this method, and an indication of which \textit{Framework(s) of Ethics} (e.g., deontological, consequentialist, virtue, pragmatist, care) were most related to the method.
\end{itemize}

We then created an alternate option for method discovery, more closely following the search filter UI design pattern used by other methods discovery websites to expand practitioners' ability to locate methods beyond solely the intentions we had created~(``Advanced Filter Page'' in Figure~\ref{fig:interactionflow}, bottom right). Users could click on one or more filter criteria and/or use a search box to filter a search results list of matching methods, which included the method name and brief description. For this page, we relied on the same method detail pages, but used the ``Search and Filter'' WordPress plug-in to allow users to filter search results for all methods based on their relevant ``cores,'' the same intentions used on the Main Intentions Page~(Figure~\ref{fig:interactionflow}, left), Audiences (e.g. industry practitioners, student/educators, industry researchers, academic researchers), Design Phases (e.g., ``Phase 1--Scoping/Defining'', ``Phase 4--Evaluation/Iteration''), Context of Use (e.g., Individual, Team), Difficulty Level (e.g., Easy, Medium, Expert), and Framework of Ethics (e.g., Consequentialist, Deontological, Pragmatist, Virtue). 

\subsection{Stage 3: Evaluation of the Website} \label{sec:stage3}
The purpose of this evaluation study was to understand how UX practitioners would experience the ethics-focused method site and how this site might relate to their goals of better supporting ethical awareness and action in their everyday technology design practices. In accordance with UX evaluation best practices, we used a multi-stage process to identify ``bugs'' in our site, and then used prioritization of these bugs to identify areas for changes to the site, followed by additional user evaluation. 

\subsubsection{Phase 1: Initial Data Collection} 
We recruited UX practitioners and students who were working or had previously worked as UX designers, interaction designers, and user researchers. To achieve a diverse pool of participants, we used a stratified sampling methodology to account for years of experience, academic experience, industry type, and the size of organization the practitioners had familiarity working in. We then selected participants for our usability testing sessions, seeking to maximize diversity across these factors while also managing practical scheduling requirements. This stratification approach provided a structured sample of participants, which provided value when examining the differences between subgroups of a larger participant pool \cite{Robinson2014-ul}. All participants provided their informed consent following the requirements of our Institutional Review Board (IRB) and were compensated \$15 USD (students) or \$25 USD (practitioners) for completing the study. The total time required to complete the study was 45 minutes or less.

\begin{table*}[!htb]
    \centering
       \caption{Participant descriptors for initial site evaluation.}
    \label{tab:participantsinitial}
    \begin{tabularx}{\textwidth}{ XXXX }
    \toprule
     \textbf{Identifier (\# yrs. exp.)} & \textbf{Professional Title} & \textbf{Professional Status} & \textbf{Task Order}\\
     \midrule
    \multicolumn{4}{c}{\textbf{Phase 1: Initial Data Collection}} \\ 
    \midrule  
    ES01 (04) & UX Designer & Practitioner & Filters >> Intentions  \\
    ES02 (02) & UX Designer & Practitioner & Intentions >> Filters \\
    ES03 (01) & UX Researcher & Practitioner & Filters >> Intentions  \\
    ES04 (0-1) & UX Designer & Student & Intentions >> Filters \\
    ES05 (0-1) & UX Designer & Student & Filters >> Intentions \\
  \midrule
    \multicolumn{4}{c}{\textbf{Phase 3: Additional Data Collection}} \\ 
    \midrule 
    ES06 (06) & UX Designer & Practitioner & --\\
    ES07 (04) & UX Designer & Practitioner & --\\
    ES08 (02) & UX Designer & Student & --\\
    ES09 (03) & UX Designer & Practitioner & --\\
    ES10 (02) & LX Designer & Practitioner & --\\
    
    \bottomrule  
    \end{tabularx}
\end{table*}

We conducted our first round of structured usability tests with five participants who had identified previous experience working in an HCI-oriented discipline, either as a full-time employee (practitioner) or through an internship (student) (Phase 1 in Table~\ref{tab:participantsinitial}). The sessions were conducted using a think-aloud protocol that allowed the participant to choose what they wanted to interact with on the website while describing what they were thinking and seeing during the interaction. All sessions took place via Zoom while using our website and were recorded, and we were able to capture all participant interactions by asking them to share their screen during the session. The structure of each session included three primary sections: familiarity with the website (1--3mins), task-oriented questions (20--30 mins), and follow-up final questions (5--10 mins). 

The task-oriented questions were framed around two key tasks and three sub-tasks, with each key task relating to a primary view of the interface: 1) the intentions-focused page (``Main Intentions Page'' in Figure~\ref{fig:interactionflow}, left); and 2) the advanced filter page (``Advanced Filter Page'' in Figure~\ref{fig:interactionflow}, bottom right). With each key task, we asked the participant: ``\textit{Think about an issue or opportunity you are facing in your industry context. [Use the filters to] find an ethics-focused method that you may want to try out in your everyday design work.}'' Within each key task, we asked the participant to: ``\textit{Identify the element or elements of the ethics-focused method that would be relevant to you or your work context}''; ``\textit{Find another method that has similar characteristics to the one you have chosen}''; and ``\textit{Find another method that has different characteristics to the one you have chosen.}'' For three participants (ES01, ES03, ES05), we began the session with the key task on the advanced filter page and then moved to the second key task with the intentions page, while for the other two participants (ES02, ES04), the order was reversed. Throughout the session, we took detailed notes during the usability testing tasks, and these notes were leveraged to ask the participant follow-up questions at the conclusion of the session.  

\subsubsection{Phase 2: Analysis and Iterative Design Changes}
All session recordings were reviewed and transcribed in qualitative analysis software. We then added first-cycle codes \cite{Saldana2015-ey} focused on: 1) structural codes relating to the tasks, 2) error-focused or sense-making codes relating to the participant's understanding of the interface, and, 3) resonance-focused codes relating to the applicability of methods information to support the practice experiences of our participants.

\begin{figure*}[!ht]
    \centering
    \includegraphics[width=\textwidth]{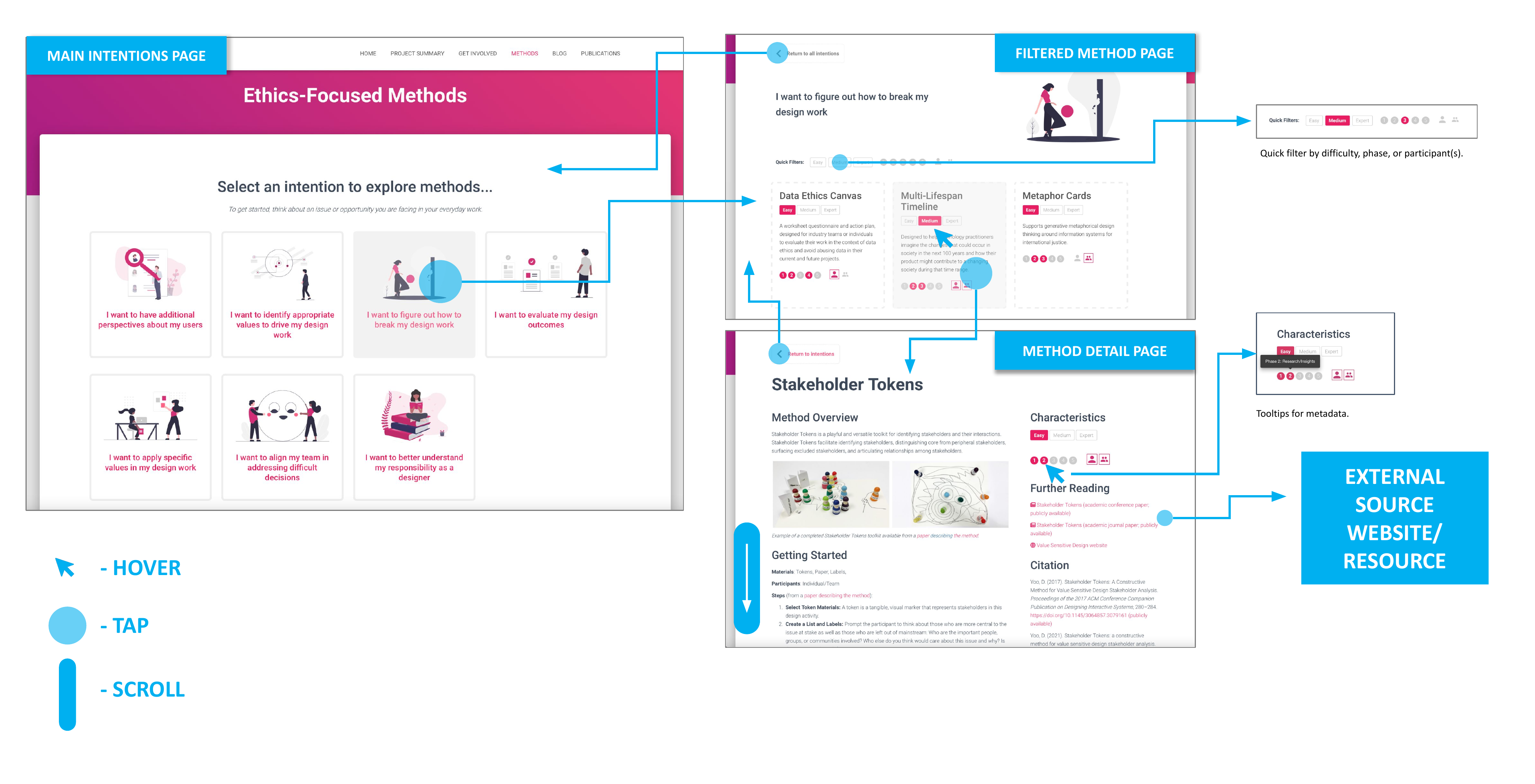}
    \caption{Revised interaction Flow for the intentions-focused navigation with additions of the quick filter and visually-focused method detail page.}
    \label{fig:interactionflowrevised}
    \Description[Interaction Flow with three images]{Interaction Flow with three images of different pages of the website. There are arrows linking clicked inputs to their connected navigation outcomes. The first image shows a page with panels depicting clickable intentions, the second image is a page of methods specific to the chosen intention with toggle-able filters that further modify the methods presented on the page, and the third image depicts an individual method page that results from selecting a method.}
\end{figure*}

Based on this initial analysis of the first five sessions, we identified a range of issues relating to participants' experience with the site that we wanted to understand, prioritize, and then use to inform changes to the site. The most common categories of concern included lack of visual details and organization, lack of description regarding the purpose or meaning of provided filters, a need for additional resources and details to situate each method, and technical bugs with filter page resetting. We then operationalized these concerns as a design team, ranking these changes based on technical feasibility, potential for impact, and time. We then made the following high-level changes:

\begin{itemize}
    \item \textbf{Removal of the Advanced Filter Page.} Our initial goals for including the advanced filter page were to establish parity with existing methods collection sites, and as a means to discover which filters were relevant to users while selecting methods. Through our initial sessions, we discovered that only a subset of filters were commonly used and appreciated by participants (i.e., phases, contexts, difficulty) and the other types of filters were either confusing or otherwise not salient (i.e., intentions, frameworks of ethics, cores). Based on this feedback, we removed the advanced filter page and the intention block linking to this page on the site, and instead incorporated relevant filtering elements into the interaction flow beginning with the intentions.
    
    \item \textbf{Addition of Quick Filters and Metadata.} Informed by the removal of the advanced filter page, we identified several key areas where participants desired additional flexibility and control in selecting methods relevant for their context that did not require reading all text or clicking through multiple methods. We added a visual indication of whether the method could be used by an individual and/or team (context) and which part(s) of a design process a method was most relevant for (phases) to the method summary block on the ``Filtered Methods Page'' (Figure~\ref{fig:interactionflowrevised}, top right). We then added a ``quick filter'' option to the top of the that page (as indicated in Figure~\ref{fig:interactionflowrevised}), allowing users to filter intention results by difficulty level (easy, medium, expert), phase (1--5), and/or context (individual, team).
    
    \item \textbf{Addition of Visually-Focused Method Descriptions.} In comparison to the text-focused Method Detail Page~(Figure~\ref{fig:interactionflow}, middle right), all participants desired a less text-heavy presentation of method details, and many participants also wanted more details that related to the application of the method in their own context while desiring fewer details that required additional scrutiny or learning (e.g., ethical paradigms). In response, we reformatted the ``Method Detail Page''~(Figure~\ref{fig:interactionflowrevised}, bottom right), adding a feature image from the method website (if available) and including other relevant multimedia resources (e.g., YouTube videos) if available. We also removed the list of intentions and ethical paradigms and added a new sidebar area that contained additional detail about the method (i.e., Further Reading, Citation). This new sidebar included the previously included difficulty rating alongside new metadata to describe the context (individual, team) and design phase(s) (1--5) in a design language identical to the intention blocks and quick filters on the previous page.
    
    \item \textbf{Iteration to Visual Organization.} In comparison to our initial flow, we observed that many participants used a lot of scrolling and unnecessary movement to get a full glance at all the information they felt they needed to understand the method or its relevance for their goals. In response, we changed the initial intention blocks to appear in two rows of four to reduce the need to scroll~(``Main Intentions Page'' in Figure~\ref{fig:interactionflowrevised}, left), and also added a caption to the main intentions page to indicate how a user could use the methods collection, inspired by our usability task language: ``\textit{To get started, think about an issue or opportunity you are facing in your everyday work.}'' We also added additional visual structure to the Method Detail Page~(Figure~\ref{fig:interactionflowrevised}, bottom right), adding icons to indicate the different kinds of resources under ``Further Reading'' (e.g., website, paper, worksheet) and indicating in text what kind of paper was being referred to (e.g., academic, practitioner); and if academic, whether it was free to access without a paywall.
\end{itemize}

These changes resulted in a streamlined task flow that removed the advanced filter flow and focused attention on navigating through intentions, followed by use of the quick filter to identify metadata that appeared to be salient to users.
 
 \subsubsection{Phase 3: Additional Data Collection and Summative Analysis}
With the streamlined version of our iterated website design complete~(Figure~\ref{fig:interactionflowrevised}), we then conducted additional usability tests to evaluate the newer version of our website. Our intentions for the usability testing remained the same as in Phase 1, but we slightly re-framed our protocol based on the edits done to the website. We simplified to a single primary task, which was now: ``\textit{Think about an issue or opportunity you are facing in your industry context. Find an ethics-focused method that you may want to try out in your everyday design work.}'' A new sub-task in the Filtered Method Page (Figure~\ref{fig:interactionflowrevised}, top right) asked: ``\textit{Now that you have selected what you need support with, use the filters to find an ethics-focused method that you may want to try out in your everyday design work.}'' We also removed the reflection question from Phase 1, asking participants to compare and contrast the intentions and advanced filter paths since we streamlined the interaction flow in the newer version. We conducted this additional evaluation with five participants (ES06--ES10) as represented under Phase 3 in Table~\ref{tab:participantsinitial}. As we conducted these sessions, we continued our analysis process using our initial codes from Phase 2 as a point of departure, noting areas of saturation in relation to usability errors and patterns of use relating to practice resonance. These codes were then grouped into themes---building on our initial analysis, design changes, and additional evaluation findings---to form the final set of heuristics described in Section~\ref{sec:heuristics}.


\section{Designer \textit{Intentions} in Relation to Ethics-Focused Methods} \label{sec:intentions}
To answer RQ \#1, we leveraged a pragmatic focus on the intentions that designers bring to design situations, focusing on how these intentions may link to the selection of appropriate tools to support ethical awareness and action. Building on prior work on designer agency as it exists alongside organizational and design complexity~\cite{valuelevers,Chivukula2021-oj,Gray2019-wa,Wong2021-pv,Steen2015-mt} and our analysis of interviews from a range of technology and design professions, we created seven ``intentions'' to frame particular kinds of expected engagement with ethics-focused methods. Below, we list, describe, and provide examples of ethics-focused methods that support each intention:

\begin{enumerate}
    \item ``\textit{I want to have additional perspectives about my users}'' describes methods that support designers in considering or better understanding user and stakeholder needs and experiences, using this knowledge to inscribe appropriate values into their design work. For example, the method GenderMag \cite{gendermag} provides step-wise instructions for software developers to evaluate existing or in-progress products with inclusive and diversified personas, aiding practitioners in considering the perspectives of users that were not considered during the development of the product. Other examples that support this intention include Stakeholder Tokens \cite{stake-tokens2}, Judgement Call the Game \cite{judgementcallthegame}, and Adversary Personas \cite{adversarypersonas}.
    
    \item ``\textit{I want to figure out how to break my design work}'' describes methods that support designers in envisioning different---and perhaps counter-intuitive or speculative---ways in which design outcomes might be understood or used in a range of use contexts. For example, the method Normative Design Scheme \cite{normativedesignscheme} provides a worksheet for designers to assess their design goal by de-structuring it using different ethical theories such as virtue ethics (to define the intention of the product), deontological ethics (to define the design), and consequentialist ethics (to achieve an effect). Other examples under this intention include Security Cards \cite{securitycards}, Envisioning Cards \cite{envisioningcards}, and Ethics Canvas \cite{ethicscanvas}.

    \item ``\textit{I want to identify appropriate values to drive my design work}'' describes methods that support designers in discovering and considering values that are relevant for their design work or specific design context. For example, the method HuValue \cite{huvalue} provides sets of value clusters and value-focused words for designers to identify appropriate set of values they would like to incorporate in their design process and product. Other examples that support this intention include Ethicography \cite{ethicography}, Values at Play \cite{Flanagan2014-hf}, Moral value Map \cite{moralvaluemap}, and Ethical Disclaimer \cite{ethicaldisclaimer}.

    \item ``\textit{I want to apply specific values in my design work}'' describes methods that provide designers with a means by which values can be linked to particular aspects of their design work or in-progress design artifacts. For example, the method Moral Agent \cite{moralagent} includes descriptions of 20 values; on each value card, a set of reflective prompts encourages the designer to generate concepts using the format ``how might the design \ldots''. Other examples that support this intention include Security Cards \cite{securitycards}, Value Sketch \cite{valuesketch}, Empathic Walkthrough \cite{empathycognitivewalkthrough}, and Design with Intent \cite{designwithintent-book}. 
 
    \item ``\textit{I want to align my team in addressing difficult decisions}'' describes methods that provide design team members with tools to build consensus and engage in conversations regarding ethical paradoxes and impacts. For example, the method Ethical Contract \cite{ethicalcontract} includes a template to clearly identify the ethical objectives of the team's design goal, responsibilities of all the stakeholders involved, and a space to place each team member's signature as a commitment to their responsibility. Other examples that support this intention include Value Dams and Flows \cite{damsandflows}, Diverse Voices \cite{diversevoices}, and The Ethical Scorecards for Design Teams \cite{ethicaldesignscorecards}.

    \item ``\textit{I want to evaluate my design outcomes}'' describes methods that provide designers with a set of evaluation criteria that encourage them to reflect on and better understand the implications of their design work. For example, the method Tarot Cards of Tech provides a set of provocations for designers to speculatively and evocatively evaluate the impact of their product using themes such as usage, scale and disruption, and equity and access. Other examples that support this intention include Ethics Canvas \cite{ethicscanvas}, Judgement Call the Game \cite{judgementcallthegame}, Layers of Effect \cite{Zhou-layersofeffect}, Moral value Map \cite{moralvaluemap}, and Inclusive Design Toolkit \cite{inclusivedesigntoolkit}.
    
    \item ``\textit{I want to better understand my responsibility as a designer}'' describes methods that provide designers with concrete ways to operationalize ethics in their everyday work. For example, the method Hippocratic Oath \cite{Zhou-hippocraticoath} provides a non-legally binding ``oath'' template which can help the designer identify and commit to a specific ethical responsibility as it relates to a value-centered design outcome. Other examples that support this intention include De-scription \cite{description}, Dichotomy Mapping \cite{Zhou-dichotomymapping}, and Multi-Lifespan Timeline \cite{multilifespand-timeline-co-design}
\end{enumerate} 
 
Each of these intentions describes a set of framing expectations that surround particular forms of expressed ethical design complexity that a designer might experience---including specific framings of ``complex and choreographed arrangements of ethical considerations''~\cite{Gray2019-wa}. Our goal in outlining these intentions is to encourage designers' engagement with tools and methods that frame ethical engagement in emancipatory and self-directed ways, building upon previous work that has focused primarily on the utility of specific methods, adherence to a disciplinary code of ethics, or frames responsibility primarily in relation to one or more dominant human values. By describing these intentions in relation to particular areas of project, team, organizational, or personal engagements with issues of ethical concern, our goal is to better connect a practitioner's awareness of ethical responsibility with the potential for ethically-focused action, building on pragmatic modes of engagement proposed by other scholars~\cite{valuelevers,VSDbook}, accounting for the felt ecological complexity of ethics in practice. These intentions, and their alignment with tools to support or extend the capacity of practitioners, may serve as a useful framework to scaffold the development of methods awareness and support the work of practitioners as ``everyday ethicists'' in their work contexts. 

\section{Heuristics to Evaluate Practice-Resonance of Scaffolds} \label{sec:heuristics}

To answer RQ \#3, we leveraged our analysis of practitioner interactions with our method website in two phases, identifying the characteristics of method discovery that appeared salient to participants and pointed towards aspects of practice resonance. In this section, we describe an initial set of six heuristics to evaluate the practice resonance of scaffolds, leveraging the design work we have already done while also serving as a foundation for the development of new practice-resonant scaffolds to support methods discovery, adaptation, and implementation.


\begin{enumerate}
    \item \textbf{Forming a Coherent Call to Action}
    Practitioners will need to interpret the ethical dilemmas they are facing in ways that link to a coherent call to action, thereby ``onboarding'' or sensitizing them to their need for additional support. This interpretation process requires the designer to anticipate and scaffold connections between felt ethical design complexity and one or more tractable paths forward. While our ``intentions'' are one way of indicating these coherent connections, other forms of intermediate-level knowledge, such as ethics-focused ``strong concepts''~\cite{Hook2012-dd}, may be useful in helping practitioners pragmatically engage with their local design context and concerns. In our website, while interacting with the Main Intentions Page, ES03 suggested ``\textit{if there's like a step before---that would be helpful, which is like: how do you even get to the point of having that problem statement to lead to the intention?}'' ES03 mentioned that the prompt that asked the participant to envision an ethical dilemma at their work made it ``easy'' and helped them feel ``confident'' to frame and select their intentions, which they otherwise felt ``\textit{might be hard for me to then go into the intentions or even go into the filters and be like, this is what I'm looking for.}'' Additionally, ES10 suggested that there could be further clarification around how the methods on the website are targeted for ``design and ethics,'' due to the repetitive use of the words ``design work,'' ``design outcomes,'' or ``design.'' Changes in this regard may help to indicate to users that the website is largely meant for ``designers.'' Practitioners [ES02, 03, 06] also desired coherence relating to the terminology used in the website as well as a method description that would be relatable to a range of practitioners. For instance, participants referenced the difficulty level and framework of ethics, reflecting: ``\textit{it could also be really hard to digest [these descriptions] for practitioners because we don't have a lot of training in academic research}'' [ES02] and ``\textit{I would have expected hard, but then there’s something called expert, providing high-level guidance such as the framework `epistemology.' I don’t know what that was}'' [ES06]. 
    
    \item \textbf{Using Action-Focused Language} 
    Practitioners benefit from the use of strong verbs, a clear order of steps, a description of where in the design process a method might ideally fit, and indications where improvisation is encouraged. As participants interacted with the method detail pages, they commonly wanted to quickly identify what they would be doing, how they would accomplish the goal set out by the method, and whether this level of effort would be ``worth it'' in advancing their design goals. These desires were supported best when the method steps were quickly understandable, supported by action-focused language that mapped with practitioners' previous design experiences. In our website, while interacting with Method Detail Page of White-Hat Design Patterns \cite{whitehat}, ES02 mentioned that having the principles listed is ``\textit{really good [\ldots] because they're all actionable things and instead of high level things to focus on. Because for high-level things, it's really hard to know what you need to do before this one.}'' Adding to this perspective, he pointed out the lack of clarity in one of the principles saying ``\textit{`be transparent'---that can be little weak.}'' ES08 pointed out the need for vocabulary in the tooltips provided on the website to be better aligned with practitioner language: ``\textit{not a lot of people, like probably only, I would say 5\% would know what these words [epistemology, framework] are.}'' We have observed that many methods with higher difficulty levels are written in ways that are more vague and abstract, and these methods in particular may benefit from more concrete operationalization into ``easy'' methods that have clearer scaffolded guidance. 
    
    \item \textbf{Evocative Exemplars or Use Cases to Demonstrate Potential Outcomes}
    Practitioners find value in seeing what a method ``looks'' like, what they could expect to complete if they invest time in using a method, or a quick onboarding guide to using a method that is less text-heavy. For example, a feedforward towards potential outcomes could be indicated by the presence of worksheets or other diagrams that might guide their use of a method, an example of a completed method outcome, or an onboarding video, inspiring practitioners and encouraging them to take the first step towards completion. In our website, while interacting with the Method Detail pages, many participants indicated the need for \textit{exemplars or case studies} to quickly understand the context of use, perhaps through ``mini scenarios'' of a method [ES01, ES03, ES07, ES08], and instances of potential repercussions if practitioners failed to use the method [ES06]; \textit{visual infographics} to illustrate the ``Steps'' to be taken to accomplish the method [ES07, ES10]; \textit{visual material} such as videos or images to know more about the method in one glance [ES02, ES04, ES05, ES08, ES09]; and \textit{visual references} of any potential tangible outcome(s) of a method as prescribed by the method developer [ES03] or as posted by method users [ES06]. For instance, ES02 (with similar thoughts shared by ES10) wanted to ``\textit{visualize the method. And I don't think [the text descriptions] help me visualize [\ldots] I really don't want to read anything at this point. I do want to see some pictures or videos that can show me how it's done}''; he felt these visuals would be ideally placed alongside the ``Steps'' listed so he ``\textit{can really see what they mean to visualize the process.}'' ES01 also asked for examples for methods that were not as straightforward or easy to understand: ``\textit{I would love for examples of when I would use something like [Adversary Personas \cite{adversarypersonas}] or when it would be useful, just so I can kind of see examples.}'' ES06 also recommended more clearly demonstrating industry value through negative use cases to share, for instance: ``\textit{this is example of some company breaking GDPR compliance somewhere. Or user backlash on some decisions that were made.}'' These kinds of use cases would help in ``\textit{raising more awareness of things that could go wrong and why businesses should be mindful.}''
    
    \item \textbf{Encouraging Many Paths}
    The diversity of practitioner experiences and goals requires the presence and support for multiple ``paths'' towards and through a method which could be accomplished by recognizing that designers might start their journey with a number of different motivations. Examples of these goals from our participants included: putting out a ``fire'' in their current practice; using a method to provide more convincing data that they are making the right decision [ES03]; ``\textit{always wanting to critically engage as  [\ldots] I want to engage designers as well as engineers to reach a common solution to a problem}'' [ES05]; educating their team or other stakeholders [ES02]; ``\textit{mov[ing] forward with my design work ethically}'' [ES02]; ``\textit{reflect[ing] on my design decisions and values}'' [ES08], communicating about ethics with stakeholders or fellow designers [ES07, ES09, ES10]; or as a part of their lifelong learning practices or method exploration [ES07, ES10]). In our website, we have provided access to external pathways to information by providing links to source website(s), which are intended to not constrain the practitioners to our website alone. However, ES10 ``\textit{was thinking maybe this page [method detail page of Microsoft Inclusive Design toolkit \cite{microsoftinclusive}] could have a little bit more about what to focus on when I go on the Microsoft website.}'' The designer of practice-resonant scaffolds might identify that these multiple paths require a range of different types of interactive ``hooks,'' some of which may deal with difficulty, while others may deal with intention, or who needs to be or can be involved.  While we already sought to provide multiple types of these ``hooks'' or paths through different interactive flows, information detail, or provision of external resources, it is also important to balance the multiplicity to avoid ``\textit{decision paralysis}''[ES05] or a ``\textit{high barrier of entry}'' [ES03], concerns which can be mitigated through a simplification of the onboarding process into the collection of methods or providing the right kind of framing detail upfront.
    
    
     \item \textbf{Providing (The Right Kind) of Detail}
    Practitioners need the right kinds of detail at the right moments in order to assess whether a method is salient for their needs and felt ethical design complexity. Most frequently, this detail is presented as method metadata, which acts as a sense-check, a tool to narrow down their options, or as a form of ``hand-holding''[ES03] to help the practitioner ascertain whether this is the right method to pursue or learn more about based on their intention. This requirement of detail is less about academic completeness (e.g., proper citations) or links to academic concepts (e.g., formal ethical paradigms) and more about performative aspects of methods use~\cite{Goodman2011-ak,Gray2022-na}. Participants suggested a range of additional metadata that might have to be added to the Filtered Methods Page to help them with selection of a method from the list such as: \textit{time} that would be spent to accomplish the method [ES01, 02, 03, 06, 07]; \textit{professional roles} that can relate to the method (e.g., ES02 mentioned ``\textit{I'm not sure if this [selected method] is something that the PM [Product Manager] can do on their own or something that designer facilitated this activity with the PMs.}'']; the \textit{focus} of the ethical conflict through a method as filters or indicators (e.g., ES08 expressed additional detail to narrow down to a method saying that all methods could have an indicator ``\textit{this is what you should do if you are having a conflict with product, or if you are having conflict with software engineer, [or] if you are having conflict making a design decision}''); \textit{information} about the medium of use and planning required to use the method [ES02, ES03]; a \textit{list} of complementary or similar method(s) to aid in exploration [ES03, ES07, ES10]; the \textit{scale} of the method relating to group size or ``piece'' of the product (e.g., ES03 expressed doubt about a method saying: ``\textit{Does that mean like the [whole] product or part of the product? [\ldots] just guidance of what size and what scale should I be doing this on would also be really helpful.}''); \textit{visual material} related to the method (as discussed in another heuristic); and \textit{additional resources} to support describing (ethics-focused) vocabulary to frame common understanding needed for the scaffolds [ES02, ES03]. 
    The designer of practice-resonant scaffolds might consider different levels of access to properties of a method that have ecological relevance, including easy access to details to support assessment of initial salience alongside links to additional resources that contain richer detail for further depth, supporting extended exploration.

    \item \textbf{Ensuring Access to (Relevant) Materials}
    Practitioners need access to relevant resources that will deepen their awareness in and potential performance of a method. However, a range of levels of resources must match practitioners' intentions and goals, without an undue focus on resources (e.g., academic papers) that are not primarily intended for a practitioner audience. Additionally, beyond the initial low perceived salience of a dense academic paper, practitioners are almost guaranteed not to read these papers if they are behind a paywall. Thus, access to all referenced materials should be freely accessible as a baseline, but ideally, practitioners would benefit more from visually-focused or practice-oriented descriptions of a method. ES05 expressed enthusiasm for the academic citations attached in the Method Detail Page as a sign of ``credibility'' that a lot of research had been done to frame a method. However, ES02 tried to access the external source website of a method that was deemed academically credible, mentioning: ``\textit{this link doesn't really provide me the information I might want to learn. I think they're not that helpful at this moment.}'' and then upon clicking the ``DOI link,'' concluded: ``\textit{the actual paper can be helpful, but the thing is, I just don't have access right now. Yeah. It's behind a paywall.}'' Additionally, practitioners desired access to the method material in one click rather than having to be re-directed to a source website and then being asked to locate and download relevant materials (e.g., for the method Ethical Design Scorecards, ES06 questioned: ``\textit{I’m assuming then download the Excel sheet from the website}''). On the method detail page of Ethical Design Scorecards \cite{ethicaldesignscorecards}, ES08 expected our website would also provide access to ``\textit{free resources}'' and was ``\textit{disappointed here because I thought I was going to get the access to research}''---namely the book that the Excel sheet was introduced in.  
\end{enumerate}

Across our iterative engagement in the design and evaluation of this method discovery site, we have built new knowledge regarding appropriate heuristics to identify weaknesses in practice-resonance, pointing towards additional design qualities that might improve the resonance of ethics-focused scaffolds. Evaluation participants provided us a range of feedback on the relevant kinds of detail that might help them to decide on the relevance of a method, but often it was difficult for us to add those details to our revised version of the website due to the lack of detail in the source method description. However, even with noted deficiencies and opportunities for further elaborated method prescriptions that provide many paths towards performance in a range of practice settings, the website---in both its initial and improved forms---was perceived as useful by all participants, with multiple participants asking if they could share the resource with other team members, considered where they might use the resource as part of their ongoing learning practices, or identified how the site might be used as a tool to build alignment in their team to address ethical issues in their work context. These heuristics have aided us in building a better appreciation for the many levels of interactivity and detail that are needed to increase practice resonance, and also provides us with a roadmap for the improvement of existing methods and design of new methods that are more likely to be perceived as practice-resonant which we will describe further in the next section.

\section{Discussion and Implications}

\subsection{Scaffolding Practice Resonance}
In this action research project, we have identified the uptakes---and also the complexity---of scaffolding existing ethics-focused methods in ways that may increase practice resonance. In some ways, the idea of scaffolding design knowledge in these particular ways is relatively new to design methods. However, both the foundational text, \textit{Universal Methods of Design}~\cite{umod}, initially published in 2012, and the proliferation of ethics, value-related, or policy-related focused toolkits provide evidence of a growing interest on the part of practitioners in supportive framings of design knowledge that consider the ethical implications of practitioners work.

In scaffolding practice resonance through this project as designer-researchers, we became aware of several different facets of scaffolding that are important to consider. First, we had to choose how to design with metadata, first identifying the characteristics of the method prescriptions we were engaging with, and then seeking to better understand which metadata elements or other forms of guidance could lay the groundwork for further means of support (i.e., the heuristic \textit{The Right Kind of Detail}). Second, we had to identify what interactional layer of content or additional guidance needed to exist in order for resources to have the potential for resonance in a range of practice-responsive setting (i.e., the heuristic \textit{Forming a Coherent Call to Action}). Third, with these two elements considered, the interplay between the content and a compelling call to action needed to be carefully crafted to respect both the agency of the practitioner and the kind(s) of difficulty that might need to be anticipated when considering application of method prescription(s) in particular contexts or in relation to specific types of ethical design complexity (i.e., heuristics \textit{Using Action-Focused Language, Encouraging Many Paths,} and \textit{Evocative Visuals to Demonstrate Potential Outcomes.}) Across the three levels that we have identified, scholarly attention has previously focused primarily on the mediation of tools in relation to the organization and the practitioner  (e.g., practitioner-focused mediation in~\cite{Wong2021-pv,Gray2019-wa} and methodologically-focused mediation in~\cite{VSDbook,Flanagan2014-hf}). However, there have also been rare instances of practitioner interest documented in the research literature that may precede or frame a call to action~\cite{Watkins2020-zr,Chivukula2020-oy,Wong2021-pv}. However, our framing of intentions as a specific and articulated interactional layer appears to be a new means of support that could be further investigated. This layer supports ethics-focused practice, including the consideration of the knowledge latent in methods that may be used to form scaffolds for method discovery, selection, and use. This finding implies the need to carefully consider resonance as it relates to the complexity of practice overall---attending to multiple facets that focus on methodological guidance (\textit{knowledge resonance}), practical limitations that can constrain or enable use (\textit{ecological resonance}), and opportunities and/or forms of knowledge that bridge prescription and performance of methods (\textit{interactional resonance})---building on Stolterman's~\cite{Stolterman2008-ho} call for ``rationality resonance'' in considering method use.

\subsection{Impacts of Practice Resonance on the Design of Ethics-Focused Methods}

Given what we were able to understand about the scaffolding of existing methods through the design and evaluation of an interactive website and set of intentions, what else have we learned about practice resonance that may positively impact the design of new forms of knowledge to support ethical awareness and action? Additionally, what opportunities might exist to redesign or augment existing methods, or generate new knowledge forms that might increase practice resonance?

First, we identified multiple areas of method insufficiency in relation to prescriptive components of existing methods. For example, we identified multiple instances where the method, as written, did not provide a clear prescriptive component that could be readily translated by practitioners into performance opportunities. This could be compared to a play script that has left out key staging instructions or lines of dialogue, even if the overall narrative arc is clear. In some instances, this lack of accessible prescriptive components was due to a primarily scholarly grounding that privileged conceptual clarity over performative actionability (e.g., the VSD method \textit{Co-Evolve Technology and Social Structure}). While in other instances, the method referred to an existing and holistic practice that must be already understood in order to engage in the present method (e.g., the \textit{Value Reflection Workbooks} method, which requires background knowledge in creating design workbooks). Other forms of method insufficiency related to the presentation layer of the method, where in general, methods that had a visible workbook or physical structure were more easily understood by practitioners. Furthermore, additional methods that either lacked any visual component, did not prioritize a visual component in describing the method prescription or potential outcomes, or lacked key logistical information (e.g., timing, suitability for various professional roles) were less likely to be perceived as resonant by a practitioner.

Second, the prescriptive opportunities embedded in the method were, at times, too robust (e.g., Microsoft's Inclusive Design Toolkit, White Hat Design Patterns). Methods of this type provided too \textit{many} points of entry which could overwhelm a practitioner rather than inspiring action as a point of departure. In these cases, there could be value in splitting up the contribution into multiple, individually-coherent methods that may lead to more flexible discovery and means of supporting performance. For instance, if \textit{White Hat Design Patterns} existed as an overarching toolkit that provided conceptual clarity and consistency, while being anchored in multiple specific methods that could be discovered separately based on different practitioner intentions. 

Third, some methods pointed towards different levels or types of performance that may rely on different levels of practitioner creativity, improvisational capacity, or ability to remix or appropriate method elements. For instance, many of the practitioner-created methods followed a predictable pattern, using worksheets that heavily structured a set of actions the practitioner should engage in. Such worksheets indicated key metadata such as timing alongside specific information they should fill in (e.g., stakeholders or unethical situations in \textit{Ethical Disclaimer}). In contrast, other methods were so improvisational that they could serve as a barrier for a practitioner who had not already built a deep capacity for reflexivity and adaptation of methods for a range of contexts (e.g., the guidance for \textit{The Oracle of Transfeminist Technologies} to ``ideate design opportunities'' and ``incorporate into your future design work'' based on open-ended reading of the cards).

These differing patterns of engagement demonstrate opportunities for both method designers and curators. Method designers might increase the prescriptive potential of methods that lack enough ``hooks'' or starting points. Additionally, they can increase the tractability of methods that are too large (i.e., toolkits) by splitting the toolkit into multiple standalone methods, and add different starting points or forms of guidance for practitioners that enter the method with different levels of creative capacity. This kind of loose structure is already present in the practitioner-created \textit{Ethics for Designers} toolkit\footnote{\url{https://www.ethicsfordesigners.com}}, for instance, which includes six distinct methods organized in relation to three key ethical themes: sensitivity, creativity, and advocacy. Similarly, method curators may need to build new kinds of actionable guidance that are not present in the original method as a means of supporting multiple levels of knowledge or potential performance, add missing metadata through testing or evaluation of methods to increase comparability, and increase the visual appeal of methods that lack such support. These kinds of actions would likely increase the practice-resonance of standalone methods and method discovery tools. However, this reorientation of method prescriptions would require a new commitment not only to method ``validity'' in the traditional performative sense~\cite{Eisenmann2021-ct}, but also a consideration of method \textit{comparability, adaptability,} and \textit{creativity} as part of a broader ecosystem of tools, as well as practitioner competencies to translate these prescriptions to shape ethically-focused practice.

\section{Conclusion}

In this paper, we outline the findings from a three-phase action research project in which we built and evaluated a tool to scaffold the discovery and use of ethics-focused methods by design and technology practitioners. Based on our creation and implementation of seven \textit{intentions}, we describe how these intentions may serve as practice-resonant calls to action that promote and support ethically-focused design work, and further outline an initial set of six \textit{heuristics of practice resonance} that may promote the evaluation and identification of appropriate scaffolds for current and future ethics-focused resources. We conclude with opportunities to address the felt ethical design complexity of design practice, using this knowledge to better articulate and inscribe practice resonance into ethics-focused methods. 

\begin{acks}
We thank graduate student Lukas Marinovic for his work on an early version of the methods website, and also gratefully acknowledge the efforts of graduate researcher Jingning Chen and undergraduate researchers Laura Makary, Matthew Will, Janna Johns, and Anne Pivonka for their contributions to collecting, evaluating, and preparing methods for inclusion. We also thank graduate researcher Aiza Hasib for her contributions to the interviews with technology and design practitioners that informed the creation of the method intentions. This work is funded in part by the National Science Foundation under Grant No. 1909714.
\end{acks}

\bibliographystyle{ACM-Reference-Format}
\bibliography{intentions}


\end{document}